
\input phyzzx

\def\Tr{{\rm Tr}}
\def\half{{\textstyle {1 \over 2}}}

\def\FIG#1{\global\advance\figurecount by 1
\xdef#1{\number\figurecount}}

\parskip=0pt plus 1pt
\def\figure#1{\midinsert\baselineskip=12pt
\Tenpoint\vskip#1\noindent}
\def\endfigure{\endinsert\par\vskip20pt}

\parindent 25pt
\overfullrule=0pt
\tolerance=10000
\def\ie{{\it i.e.}}

\def\half{{\textstyle {1 \over 2}}}

\def\Tr{{\rm Tr}}

\sequentialequations
\nopagenumbers
 \baselineskip=15pt

\hfill {QMW-92-18}\break
\null\hfill {NI92010}\break
 \null
\vskip 2cm
\centerline{THE INFLUENCE OF WORLD-SHEET BOUNDARIES}
\centerline{ON CRITICAL CLOSED STRING THEORY}
\bigskip
\bigskip
 \centerline{ Michael B.  Green,}\foot{email: MBG@V1.PH.QMW.AC.UK}
 \smallskip
 \centerline{Department of Physics,  Queen Mary and Westfield College,}
\centerline{Mile End Road, London E1 4NS, UK}
\smallskip
\centerline{and}
\smallskip
\centerline{Isaac Newton Institute,}
\centerline{20 Clarkson Road, Cambridge CB3 0EH, UK}
\vskip 1.5cm
\nopagenumbers
\bigskip
{\centerline { ABSTRACT}}\hfil\break\noindent
This paper considers interactions between closed  strings and open
strings  satisfying either Neumann or constant (point-like)
Dirichlet  boundary conditions in a BRST formalism in the critical dimension.
With Neumann conditions
this reproduces the well-known stringy version of the Higgs mechanism.
With Dirichlet conditions the open-string
states correspond to either auxiliary or Lagrange multiplier
target-space fields and their coupling to the
closed-string sector leads to constraints on
the closed-string spectrum.

\vfill\eject

\pagenumbers
\pagenumber=1

Conventional closed-string theory is formulated in terms
of  a sum over world-sheets
with no boundaries.  Adding boundaries in the usual manner,
\ie\ with Neumann boundary conditions on the space-time
coordinates, $X^\mu(\sigma,\tau)$, leads to a theory with
both closed and open
strings with free end-points.  Adding boundaries with constant
Dirichlet conditions  leads to a
theory with no physical open
strings but with a radical modification of the
closed-string theory (in which, for example, fixed-angle
scattering behaves in a point-like manner).   With orientable
world-sheets (which is all that is considered here) the boundaries
carry quantum numbers of the defining representation of a
unitary group, denoted $U(m)$ (or \lq flavour')  in the Neumann case
and $U(n)$ (or \lq colour') in the Dirichlet case.

\FIG\fone{}
\figure{1.17in}
 \smallskip Figure~\fone.\
(a)    A representation of  the interaction of a number of
on-shell closed-string states on a
world-sheet with the topology of a disk.   The thick line indicates the
world-sheet boundary.
(b)  The same process in a configuration in which the boundary represents a
closed string disappearing into the
vacuum.   With Neumann conditions the boundary couples to a
linear combination of the dilaton and
the trace of the graviton in the cylinder while with
Dirichlet conditions only the dilaton couples
to the boundary.
(c)  Illustration of an intermediate open string.  With Dirichlet conditions
this
string has both end-points fixed at the same space-time
point.
\endfigure
Earlier work (see \REF\greenb{M.B. Green,
{\it Space-time duality and Dirichlet string theory},
Phys.  Lett. {\bf 266B} (1991) 325;   {\it Point-like Structure in String
Theory},  talk
presented at the First
International Sakharov Conference, Moscow (27-31 May,
1991).}[\greenb] and
references therein) has pointed out certain special features
of the Dirichlet
theory and its relationship, via space-time duality, to the
Neumann theory.   A world-sheet with a single boundary (the
disk illustrated in fig.1(a)) may
be parametrized so that the boundary is the end-state of a
cylindrical section of world-sheet
(fig.1(b)).
The process may then be expressed in terms of  the evolution
of a
closed string coupling to the boundary,
which is represented by a state, $\langle\langle B|$ (the double ket notation
indicates a state  defined in the product of the spaces of left and
right-moving modes), where
$\langle \langle B |\partial_n X^\mu  =0$  in the
Neumann theory (the subscript $n$ indicates a derivative
normal to the boundary), whereas in the
Dirichlet theory $\langle \langle B | (X^\mu - y_B^\mu)=0$, where the
position of each boundary must be integrated (and the total momentum passing
through
the boundary vanishes as a result).
\REF\callana{C.G. Callan, C. Lovelace, C.R.
Nappi and S.A. Yost,
{\it Adding holes and cross-caps to the superstring},
Nucl. Phys. {\bf B293} (1987) 83.}\REF\polchina{J.
Polchinski and Y. Cai, {\it
Consistency
of open superstring theories}, Nucl.  Phys. {\bf B296}
(1988) 91.}The end-state bra, given in terms of
oscillator states by
[\callana,\polchina,\greenb]
$$\langle \langle  B |= \langle  \langle \uparrow \uparrow| ( B_0 - \tilde
B_0)\exp\sum_{n=1}^\infty (\pm {\alpha_{-n}\cdot \tilde
\alpha_{-
n}\over n} -C_{-n} \tilde B_{-n} -
\tilde C_{-n} B_{-n}) , \eqn\boundary$$
satisfies $\langle\langle B| Q_c=0$ (where $Q_c \equiv Q + \tilde Q$ is the
closed-string
BRST operator).  The plus sign corresponds to the Neumann theory and
the minus sign to the Dirichlet theory (whereas the ghost coordinate boundary
conditions are
determined by the world-sheet geometry, independent of the boundary conditions
of the space-time
coordinates).
The harmonic oscillators $\alpha_n^\mu , \tilde
\alpha_n^\mu$ are the modes of the left-moving and
right-moving closed-string embedding coordinates and $B_n,
\tilde B_n,C_n,\tilde C_n$ are the modes
of the antighost and ghost coordinates.  The ground state
$\langle \langle\uparrow \uparrow|$ (${\equiv \langle \langle \downarrow
\downarrow| C_0\tilde C_0 }
 $) is the product of the ghost-number $\half$ ground states
in the left-moving and right-moving  closed-string space of states which is
annihilated by $C_0, \tilde C_0$ and has zero momentum
(which follows in the Dirichlet case after integration over
the boundary position, $y_B^\mu$).   The closed-string
propagator contains a factor of  $(C_0-\tilde C_0)$  so that \boundary\ implies
in the
Neumann case that a
linear combination of the trace of the graviton and the
dilaton propagate in the cylinder.
This leads to a divergence that signals the breakdown of conformal invariance
from a
boundary of moduli space that may be compensated by
a change in the background fields that
introduces a cosmological term in the target-space theory
\REF\fischlera{W. Fischler and L. Susskind, {\it Dilaton tadpoles, string
condensates and
scale invariance},
{ Phys. Lett.} {\bf 171B} (1986) 383; {\it Dilaton tadpoles, string condensates
and
scale invariance. II}, { Phys. Lett.} {\bf
173B} (1986) 262.}[\fischlera].  By contrast, in the Dirichlet
theory the only massless state that propagates in the cylinder is the dilaton
causing a
zero-momentum divergence
that will be addressed later.
The term \lq dilaton'  is here reserved for the massless
general coordinate scalar state.

The disk amplitude in fig.1(a) can also be expressed as
a sum over intermediate open strings,
as illustrated in fig.1(c).   In the Neumann theory this
leads to the mixing of open-string
and closed-string particle states but in the Dirichlet
theory the presence of intermediate open
strings leads to further divergences.  This problem (noted and
discussed in [\greenb]) arises because the intermediate open
string in
fig.1(c) has both end-points fixed at the same
space-time point.  The level-one vector state, in particular,
should be identified with a space-time Lagrange multiplier
field.  The divergences arise in string
perturbation theory due to the fact that such a field has no
mass or  kinetic terms  and hence has a singular propagator.
Obviously, perturbation theory is inadequate in this
situation and the presence of such  fields is a signal that
constraints should be imposed on the closed-string states to
which
they couple.  One purpose of this paper is to
elucidate these constraints in more detail.

This  requires a discussion of
the cohomology of the BRST operator in the open-string
sector of the Dirichlet theory together with a description
of the coupling of  open strings to a Dirichlet boundary.
It will prove useful to compare the Dirichlet case with the
more familiar case of Neumann boundary conditions.

\vskip 0.3cm
\noindent{\it BRST cohomology and free string fields.}
\par\vskip 0.1cm\nobreak
Physical states of  Neumann strings (open strings with free
end-points)   represent particles of definite mass (such as
the massless photon)  with wave-fuctions that satisfy
equations of motion when the BRST constraints,
$Q_o|\Psi\rangle =0$,  are imposed (where
$Q_o$ is the open-string BRST operator which will be
denoted by $Q_N$ and $Q_D$ in the Neumann and Dirichlet
cases, respectively). These constraints may be interpreted as the free-field
equations for the
open-string field, $|\Psi\rangle$ (which is an $m\times m$ hermitian matrix).
The physical
states have ghost-number $-\half$ and are defined up to gauge transformations,
$\delta |\Psi\rangle
= Q_o|\epsilon\rangle$
(where $|\epsilon\rangle$ is an arbitrary ghost-number $-3/2$ state).
For example, the massless states
at level 1 satisfy
$$Q_N \left(A_\mu \beta^\mu_{-1} + \omega c_0 b_{-
1}\right) |\downarrow\rangle =0, \eqn\brsopen$$
where $|\downarrow\rangle$ is the ghost-number $-\half$
state annihilated by all the positive modes of the open-string conformal
gauge antighost and ghost coordinates, $b_n$,$c_n$ (with
positive $n$), as well as by $b_0$ (and the open-string modes are denoted
$\beta_n$ in order to
distinguish them from the closed-string modes).  The gauge transformations that
relate these states
are associated with the gauge parameter $\lambda b_{-1}|\downarrow\rangle$
(which is the massless
component of the general ghost-number $-3/2$ gauge parameter).   Equation
\brsopen\
requires $\omega=  ik_\mu A^\mu $ and $k^2 A_\mu +i k_\mu
\omega=0$ and Maxwell's equations result upon eliminating
the auxiliary field $\omega$.  These equations follow from
a target-space field theory action principle based on the action,
$$S_0^\prime = \int d^Dx \left(-\half \omega^2 + i\omega k_\mu A^\mu -
\half A^\mu k^2 A_\mu\right)  \eqn\vecact$$
(where $k\equiv -i\partial/\partial x$ and the subscript $0$ indicates that the
action refers to the
massless, or low-energy, sector of the complete string theory), which
illustrates
the characteristic feature of auxiliary fields that their
action does not
contain kinetic terms.  The cohomology of
$Q_N$ also contains the $SU(1,1)$-invariant ground state
which is a
scalar of ghost-number $-3/2$ and may be represented by
$\Lambda b_{-1} |\downarrow\rangle$.  This is an isolated
state since  the BRST constraint requires $k^\mu \Lambda=0$.
The
remaining non-trivial cohomology classes are the duals of
the above states,
with ghost-numbers $1/2$ and $3/2$.

World-sheets with Dirichlet boundaries  are   correlation
functions of boundaries at different space-time points.
They can be
constructed by sewing together open-string vertices with
propagators that
describe the world-sheet evolution of a string with
end-points fixed in
space-time.  The propagator for such a string is given by
$\Delta_D = (L_{0D} -1)^{-1}$,   where (in units in which
$\alpha'=1$)
$$L_{0D} = \left({y_2-y_1\over 2\pi}\right)^2  + N
\eqn\dirprop$$
and $N$ is the level number for the states of a string with
end-points fixed at space-time points $y_1^\mu$ and
$y_2^\mu$.
The singularities of $\Delta_D$ determine the space-time
singularity
structure of the correlation functions [\greenb]. The BRST cohomology
of these open-string states is
isomorphic to those of the Neumann theory with the momentum
$k^\mu$ in $Q_N$  replaced by the separation $(y_2-y_1)^\mu/2\pi$  in $Q_D$.
This
relationship between the Neumann and Dirichlet theories is a
manifestation of space-time duality in the context of open string theory.

The gauge-singlet Neumann open-string states mix with
closed-string states \REF\cremmera{E.
Cremmer and J. Scherk, {\it Spontaneous dynamical breaking
of gauge symmetry in dual models}, Nucl.  Phys.  {\bf B72}
(1974) 117.}\REF\kalba{M. Kalb and P. Ramond, {\it Classical
direct interstring action}, Phys.  Rev. {\bf D9} (1974)
2273.}\REF\shapiroa{J.  Shapiro and C.B. Thorn,
{\it BRST-invariant
transitions between closed and open strings},
Phys. Rev. {\bf D36} (1987) 432.}  [\cremmera-\shapiroa]
giving rise to a stringy version of the
Higgs mechanism that will be considered in more detail below.
In particular, the massless closed-string
antisymmetric tensor  mixes with the massless  level-one
open-string vector and gains a mass of order $g^2$ (where $g$
is the open-string coupling constant and $g^2$ is
proportional to the closed-string coupling constant).  The effect
of this mixing can also be seen by considering the coupling of a background
electromagnetic
potential to an end-state \REF\calland{A.  Abouelsaood,
C.G. Callan, C. Lovelace, C.R.
Nappi and S.A. Yost, {\it Open strings in background gauge fields }, Nucl.
Phys.
{\bf B280} [FS18] (1987) 599.}\REF\callanc{C.G. Callan, C. Lovelace, C.R.
Nappi and S.A. Yost,
{\it Loop corrections to superstring equations of motion},
 Nucl.Phys. {\bf B308} (1988) 221.}\REF\zwiebachb{B.  Zwiebach, {\it
Quantum open string theory with manifest closed-string factorization}, Phys.
Lett.
{\bf 256B} (1992) 22.}[\calland-\zwiebachb].
 In the corresponding discussion of the mixing of
 closed
and open strings in the Dirichlet theory we will see that
a special r\^ole is  played by  gauge-singlet
Dirichlet open-string states with $y_1=y_2=y$
-- \ie\ states in which the string end-points are fixed at
the {\it same} space-time point.    For every value of $y^\mu$ the BRST
cohomology
is the same as that of the $k^\mu=0$
states of the Neumann theory,  even though their wave
functions depend on $y^\mu$ and therefore do
not have zero momentum.     The only BRST non-trivial
physical state at ghost-number $-\half$ is the level-one
vector, $A^\mu$.  Since it is physical without the need
for any constraint it is interpreted in field theory as a
Lagrange-multiplier field -- it does not appear in the
free-field action but enters only via its coupling to the
closed-string sector.  The wave function of the other
level-one ghost-number $-\half$ state remains auxiliary but
now vanishes ($\omega=0$),  which is reproduced by the free
field action  $S_0' = \int d^Dx \half \omega^2$.
The constraint on the ghost-number $-3/2$ wave function,
$\Lambda$, is no longer
required for  it to be BRST non-trivial.  Its wave function
is therefore promoted from a constant  to the status of a
Lagrange multiplier field.

Before discussing the interaction between open and closed
strings in more detail I will briefly review the closed-string
sector.  In current formulations of closed-string field theory\foot{For
the latest word on this subject and an up-to-date
review of the literature see \REF\zwiebacha{B.  Zwiebach,
{\it Closed string field theory: quantum action and the B--V master equation},
IASSNS-HEP-92/41, MIT-CTP-2102.}
[\zwiebacha].  The discussion in this paper will not involve  interactions
between closed strings.}  it is necessary
to restrict the closed-string field, $|\Phi\rangle\rangle$,
so that
$$B_0^-|\Phi\rangle \rangle = 0, \qquad \qquad L_0^-  |\Phi\rangle\rangle = 0,
\eqn\closedhom$$
where $B_0^\pm  \equiv  (B_0   \pm \tilde
B_0)/\sqrt 2$ and  $L_0^\pm  \equiv  (L_0  \pm \tilde
L_0)/\sqrt 2$ are the zero-mode Virasoro generators.
Physical states satisfying \closedhom\ are \lq semi-relative' cohomology
classes of $Q_c$ (\ie\ they also satisfy $Q_c |\Phi\rangle \rangle
=0$)  and have  total ghost-number $-1$.  The free theory is invariant under
the gauge
transformations $\delta |\Phi\rangle\rangle =  Q_c |
\Xi\rangle\rangle$, where $|\Xi\rangle\rangle$ is an
arbitrary
ghost-number $-2$  state satisfying $L^-_0
|\Xi\rangle\rangle= B_0^- |\Xi\rangle\rangle =0$.  The massless
states
(level one in both the left-moving and right-moving sectors)
consist  of  a second-rank tensor, two vectors and two
scalars,
$$\left(h_{\mu\nu} \alpha_{-1}^\mu\tilde \alpha_{-1}^\nu
+ \eta_\mu \alpha_{-1}^\mu C_0^+ \tilde B_{-1}
+  \tilde
\eta_\mu \tilde \alpha_{-1}^\mu  C_0^+  B_{-1}
+ \phi
\tilde C_{-1} B_{-1} + \tilde \phi   C_{-1} \tilde B_{-1}
\right)|\downarrow \downarrow\rangle\rangle .\eqn\levone$$
Correspondingly the massless gauge invariances are defined by the ghost-number
$-2$ state
$(\xi_\mu \alpha_{-1}^\mu \tilde B_{-1} + \tilde \xi_\mu \tilde
\alpha_{-1} ^\mu  B_{-1} +  \rho B_{-1} \tilde B_{-1} C_0^+ ) |\downarrow
\downarrow\rangle\rangle $,
which contains one scalar  and two vector gauge parameters.
The dilaton field  is the combination
$D= h^\mu_{\ \mu} - \phi
-\tilde \phi$,which is invariant under all these transformations.

The massless physical closed-string states satisfy free equations of motion
that may be derived
from the gauge-invariant free-field action,
$$\eqalign{ S_0^{\prime\prime}= &\int d^D x \left( -\half h_{\mu\nu} k^2
h^{\mu\nu}   +\phi k^2\tilde \phi  -\half (k^\nu h_{\mu\nu} - k_\mu \tilde
\phi)^2\right.\cr &
\left. -
\half (k^\nu h_{\nu \mu} - k_\mu  \phi)^2 +\half
(i\eta_\mu -  k^\nu h_{\mu\nu} + k_\mu \tilde \phi)^2 + \half
(i\tilde \eta_\mu -  k^\nu h_{\nu\mu} + k_\mu \phi)^2\right)
.\cr}\eqn\statesclosed$$
The parentheses involving the vector auxiliary fields $\eta$
and $\tilde \eta$  are gauge invariant and are eliminated by
shifting these fields and integrating them, whereupon the
combination
$(\phi-\tilde \phi)$ decouples.  The remaining
terms make up the usual linearized action for  a second-rank tensor
$h_{\mu\nu}$ interacting with a dilaton  $D$.

\vskip 0.3cm
\noindent{\it The coupling between open and closed strings.}
\par\vskip 0.1cm\nobreak
String field theory provides a succinct framework for describing the effect of
the  interactions
between open and closed strings.  The free-field action can be written in the
form \REF\neveua{A.
 Neveu, H.  Nicolai and P.  West, {\it New symmetries and ghost structure of
covariant string
 theories}, Phys.  Lett.  {\bf 167B} (1986) 307.}\REF\wittena{E.
 Witten, {\it Non-commutative geometry and string field theory}, Nucl.  Phys.
{\bf B268} (1986)
 253.}[\neveua,\wittena],
 $$ S^{(0)}   =\half \int \left(\Phi C_0^- Q_c \Phi + \Psi Q_o \Psi\right)
\equiv
 \half\int d^Dx \left( \langle\langle \Phi| C_0^- Q_c|\Phi\rangle\rangle +
\Tr\langle  \Psi
	  | Q_o|\Psi\rangle \right), \eqn\freefield$$
where $\langle \Psi | =|\Psi\rangle^\dagger$ and $\langle\langle \Phi |
=-|\Phi\rangle\rangle^\dagger$
[\zwiebacha].
The open-string field  is a matrix, $|\Psi\rangle^A_{\ B}$ ($A,B = 1,\dots,M$),
in the $M\times M$-dimensional representation  of the Lie algebra
of $U(M)$  and
$\Tr$ denotes the trace in this space.  The  closed-string term and the Neumann
theory open-string
term (\ie\ with $Q_o=Q_N$ and $M=m$) reduce  to  sums of
infinite numbers of massive free-field actions when expanded in
component fields. The Dirichlet theory open-string kinetic term (\ie\ with
$Q_o=Q_D$ and $M=n$)
gives an infinite set of mass terms, one for each auxiliary component field.

\FIG\ftwo{}
\figure{1.17in}
\smallskip Figure~\ftwo.\
(a)   The coupling of a closed string to an open string in
the \lq light-cone-like frame' in which the vertex is simply
the
overlap of the closed-string and open-string coordinates.
(b)  The same coupling in the \lq vertex-operator frame', in
which the process is described by a local  vertex for the
emission of an arbitrary open-string state from a point on
the boundary end-state of the closed string.
\endfigure
 The vertex for the $O(g)$ coupling of an arbitrary  closed-string
state to an open-string state has been  described in several
ways. \REF\kakua{M.  Kaku and K.
Kikkawa, {\it Field theory of relativistic strings.  II Loops and pomerons},
Phys.  Rev. {\bf D10}
(1974)
1823.}\REF\greend{M.B.  Green and
J.H.  Schwarz, {\it Field Theory of Superstrings}, Nucl.
Phys. {\bf B243} (1984) 475.}  It was deduced in the Neumann
theory in terms of light-cone gauge field theory (where the
Fadeev--Popov  ghost contribution is absent) in [\kakua,\greend]  and
in a \lq light-cone-like' ($lcl$) frame in [\shapiroa] (where the ghosts
were included).  This is the frame illustrated in fig.2(a),
in which the incoming closed string and the
outgoing open string have the same width and the interaction
is described simply by the change of boundary conditions at
the interaction time.   The vertex had originally been
deduced by
factorization of open-string loop amplitudes
\REF\cremmerb{E. Cremmer, J. Scherk, {\it Factorization of the pomeron sector
and
currents in the dual resonance model}, Nucl.
Phys. {\bf B50 }(1972) 222 . }\REF\clavellia{J. Clavelli,
J. Shapiro, {\it Pomeron factorization in general dual models}, Nucl. Phys.
{\bf B57
} (1973) 490 . }  [\cremmerb,\clavellia]  which gave an
expression appropriate
to the \lq vertex-operator' ($vo$) frame (but without consideration of ghosts)
illustrated in figure 2(b).
The generalization to the Dirichlet theory in both these
frames (as well as the inclusion of ghost factors in the vertex-operator frame)
is given in
detail in \REF\greene{M.B.  Green and P. Wai,  {\it
Inserting boundaries in world-sheets},
QMW preprint (in preparation).} [\greene].  [The discussion could equally be
given in a
frame  relevant to Witten's formulation of open-string field theory [\wittena],
coupled to
closed strings in the manner suggested in [\zwiebachb].]

The vertex will be denoted $\langle \langle\langle V|$  (where  the triple bra
indicates the product of the states of the closed string and open
string),  with superscripts $^{lcl}$ or $^{vo}$ distinguishing the frames and
subscripts
$_N$ and $_D$ distinguishing the boundary conditions where appropriate. It can
be
expressed in the form ${ \langle\langle\langle V| = \langle
\langle\langle\uparrow \uparrow;
\uparrow |\delta^A_{\ B} \exp \Delta }$,
where $\Delta$ is a bilinear form in the annihilation
modes of the closed-string  and open-string  spaces and the state
$\langle\langle\langle \uparrow
\uparrow;
\uparrow |$ is the ground state of
ghost-number $3/2$ (the arrows successively label the states of the ghost
zero modes in the  spaces
of the left-moving and right-moving closed-string modes  and the open-string
modes).
The structure of the vertex guarantees the
continuity of the space-time and ghost coordinates and their
conjugate momenta at the interaction time
(apart from an anomaly in the conservation of the antighost  coordinates that
is
required in order that the total ghost number changes by $3/2$ due to the
boundary curvature at the
interaction point).  This in turn implies the BRST condition
$$ \langle\langle\langle V|   (Q_{c} + Q_o)  =0 \eqn\brsvert$$
(where $Q_o = Q_N$ or $Q_D$, depending on the boundary
conditions).
The explicit expressions for the vertices in $lcl$ or $vo$ frames the Neumann
and Dirichlet
theories are given in [\shapiroa] and  [\greene].

The vertex  not only determines the interaction between open and closed strings
but the
free boundary state \boundary\ can also be obtained
by coupling the open-string $SU(1,1)$-invariant vacuum state to the vertex, so
that in the
$vo$ frame,
$$\langle\langle B|
=\langle\langle \langle V^{vo}| b_{-1} |\downarrow\rangle.\eqn\freebee$$

Although the constraint \brsvert\ is an essential element in the BRST
invariance of
scattering amplitudes on world-sheets with boundaries it is
not the whole story.
Generally there are anomalies in the free-field  BRST
symmetry associated with the boundary interaction
which  are compensated by a new
transformation of the string fields.

The interaction  can be written in the form
$$  S^{(1)} = g\int d^Dx \langle\langle\langle V| \Phi\rangle\rangle|
\Psi\rangle   \eqn\fieldint$$
and the combined action $S=S^{(0)} + S^{(1)}$ is the part of the full action
that is bilinear
in string fields.  It is invariant at  $O(g)$ under the gauge
transformations,
$$\delta |\Psi\rangle = Q_o |\epsilon\rangle - g\langle
\langle\Xi|V\rangle\rangle\rangle, \quad
\delta |\Phi\rangle\rangle = Q_c |\Xi\rangle\rangle - g \langle \epsilon| B^-
|V
\rangle\rangle\rangle.\eqn\gaugesymm$$
The invariance of the action at this order is verified with the help of
\brsvert\ together with the closed-string field constraints,  \closedhom.
The modification of
the free-field symmetries
induced by the boundaries is evident from the $O(g)$ terms in \gaugesymm.
These terms mix the
transformations of the states at all levels in a complicated manner.

The action $S^{(0)} + S^{(1)}$ generates tree diagrams that mix open and closed
strings.
The complete action should also include the cubic open-string interaction (of
order $g$) as well as
the cubic and higher-order
interactions between closed strings -- which  lead to non-linear modifications
of the gauge
symmetries.   There is no consistent way of separating trees from open-string
loop diagrams that
contribute at the same order in $g$ and such loops are crucial in verifying
that the
linearized transformations
\gaugesymm\ are symmetries of the complete action.  In more detail, applying
the
transformations \gaugesymm\  to the action $S$ results in two $O(g^2)$ terms.
One of these is
proportional to  $\int g^2  \langle\langle\langle V|\Xi\rangle\rangle
\langle\langle\Phi| V' \rangle\rangle\rangle$,
where the $^\prime$ indicates a  second set of open and closed-string spaces
and the two vertices are contracted on the open-string space to form
a world-sheet with the topology of fig.1(c).  This term  vanishes
due to the fact that both $\langle\langle \Phi |$ and $ |\Xi\rangle\rangle $
are proportional
to $B_0^-$ and that $B^-(\sigma)$ is continuous at each vertex.  The second
$O(g^2)$ term in the
variation of $S$ is proportional to $\int  g^2  \langle\langle\langle V|
\epsilon
\rangle B_0^- \langle \Psi |V'\rangle\rangle\rangle$, in which two vertices are
contracted on
the closed-string states so that the world-sheet has two boundaries and the
topology of a cylinder
with $|\epsilon\rangle$ and $\langle \Psi|$ attached to separate boundaries.
This term
is non-zero but is cancelled by a
quantum open-string loop correction that arises from the double contraction of
two cubic
open-string vertices.   The cancellation of  terms in the variation of the
complete action that are
quadratic or of  higher order in the fields will not be considered here.

In principle, the $O(g^2)$ mass shifts could be calculated by integrating out
the open-string
fields, resulting in the effective
quadratic action for $|\Phi\rangle\rangle$ (in the
Siegel gauge $b_0 |\Psi\rangle =0$),
$$S_{\Phi} = \half \int d^D x
\langle\langle\Phi|\left(C_0^-Q_c - g^2M\langle r|V\rangle\rangle\rangle
{b_0\over
L^o_0-1} \langle\langle\langle V'|r\rangle
\right)|\Phi\rangle\rangle\eqn\effphi$$
(where $L_0^o$ is the zero-mode Virasoro generator in the open-string sector
and
the sum over the complete set of states, labelled $r$, contracts the vertices
on the
open-string space).   The modified closed-string masses should correspond to
the zero eigenvalues
of the inverse $|\Phi\rangle\rangle$ propagator,
$\left( L_0^+ - 2 - g^2M\langle r|V\rangle\rangle\rangle {b_0\over
L^o_0-1} \langle\langle\langle V'|r\rangle B_0^-B_0^+\right)$ (in the
gauge  $B_0^+|\Phi\rangle\rangle =0$).  However, in the
Dirichlet theory (where $L^o_0 = L_{0D}$) $(L_0^o -1)^{-1}$
is singular due to the $N=1$ vector Lagrange-multiplier field,
which implies constraints on the closed-string field that require separate
consideration.   The resulting low-energy spectrum will now be considered
separately in the
Neumann and Dirichlet cases since they are so  different.

i)  {\it The Neumann case}\hfill\break
In this case only the massless states in both the open and closed-string
sectors survive the
low-energy limit and result in a self-contained gauge-invariant field theory.
In the
light-cone-like frame (considered in [\shapiroa]) the vertex has no momentum
dependence  so there is no direct
coupling of $A^\mu$  to $h^{\mu\nu}$  and the interaction \fieldint\  couples
the gauge-singlet vector field  to the auxiliary fields ($\eta^\mu,
\tilde\eta^\mu,\phi,\tilde \phi$),
$$S_0^{(1)} = - g\sqrt m \int d^Dx \left(\hat A^\mu
(\eta-\tilde \eta)_\mu +  \hat \omega (\phi-\tilde \phi)\right).
\eqn\interboun$$
 where $\hat A^\mu \equiv \Tr A^\mu/\sqrt m$ and $ \hat\omega \equiv
\Tr\omega/\sqrt m$.
 The higher-mass string states contribute
to effective mass terms for certain massless fields, resulting in
$S_0^{(2)} = g^2 m\int d^D x \left(-  h^{[\mu\nu]} h_{[\mu\nu]} +\half (\phi
-\tilde \phi)^2 -
 \hat A^2\right)$.
The complete action $S_0^{\prime} + S_0^{\prime\prime} + S_0^{(1)} + S_0^{(2)}$
is invariant under
the \lq massless' gauge transformations,
$$\eqalign{&\delta h^{ \mu\nu } =   ik^\mu \tilde \xi^\nu +i k^\nu  \xi^\mu,
\qquad \delta\phi =i k_\mu\xi^\mu + \rho + g \Tr\lambda,  \qquad
\delta\tilde \phi = ik_\mu\tilde \xi^\mu -\rho - g \Tr \lambda \cr &
\delta \eta^\mu = k^2 \xi^\mu - ik^\mu \rho - g^2m(\xi^\mu - \tilde \xi^\mu),
\qquad\delta\tilde\eta^
\mu=k^2\tilde\xi^\mu+i k^\mu\rho+g^2m(\xi^\mu - \tilde \xi^\mu)\cr &
\delta A^\mu = ik^\mu \lambda + g
(\xi^\mu - \tilde \xi^\mu)\delta^A_{\ B},
\qquad \delta \omega=-k^2\lambda -2g\rho\delta^A_{\ B} - 2g^2\lambda, \cr}
\eqn\ordergg$$
where the $O(g)$ terms are just the  massless components of \gaugesymm\ and the
$O(g^2)$ terms
are induced by  the  massive fields in the functional integral.
  Upon integrating out the auxiliary fields the part of the resulting effective
low-energy
  linearized action  involving  $h^{[\mu\nu]}$ and $\hat A^\mu$ is
$$\int d^Dx\left( {3\over 4}  k_{[\mu}h_{\nu\rho]}k^{[\mu}h^{\nu\rho]}- (g\sqrt
m
h_{[\mu\nu]}-ik_{[\mu}
\hat A_{\nu]})(g\sqrt m h^{[\mu\nu]}- ik^{[\mu}\hat A^{\nu]})\right),
\eqn\higgs$$
which is the gauge-invariant Higgs-like action describing a
massive antisymmetric tensor field.
The action for the other fields, $h^{(\mu\nu)}$ and $D$ remains unaltered.

In the vertex-operator frame
(as originally described in [\cremmera])
the boundary simply gives rise  to a direct coupling between the vector field,
$\hat A^\mu$,
and the antisymmetric tensor field,  $h^{[\mu\nu]}$,
of the form $g\sqrt m \hat A_\mu \partial_\nu h^{[\mu\nu]}$ (where $g$ is
the open-string coupling constant).  A mass term, $ g^2 m
h_{[\mu\nu]}h^{[\mu\nu]}$ is again
generated,  leading once more to the action \higgs\ for $h^{[\mu\nu]}$ and
$\hat A^\mu$  when the  auxiliary fields are integrated.

(ii) {\it The Dirichlet case}\hfill\break
The systematics of the Dirichlet theory are quite different since it is not
possible to
argue that
the states of the open strings with identified end-points (the auxiliary and
Lagrange multiplier
fields)  decouple at low energies, even though
the massive closed-string states do.
However,  the divergent perturbation theory contributions due to the coupling
of the Lagrange multiplier field $A^\mu$
to closed strings, $S_A^{(1)}$,
may be dealt with by isolating this term in the interaction.

The  vertex-operator frame for the Dirichlet theory
has a particularly simple physical interpretation,  in which the process in
fig.2(b) is described as the amplitude for
a closed string to collapse to a single space-time point at the  interaction
time, whereupon it
transforms
into the open string.  In this frame the interaction
of the level-one vector Lagrange multiplier field  with the boundary state has
the expected
simple  form of a vertex operator,
$$\eqalign{S_A^{(1)vo}  & = g\sqrt n \int d^Dx
\langle\langle\langle V^{vo}  |\Phi \rangle\rangle\beta_{-1}^\mu
|\downarrow\rangle  A_\mu (x) \cr
& =  g \sqrt n\int d^D x\langle \langle B |\oint d\sigma_B  (C(\sigma_B)
\partial X^\mu(\sigma_B) - \tilde C(\sigma_B)
\bar \partial X^\mu(\sigma_B))|\Phi\rangle\rangle  \hat A_\mu(x) \cr
&= g\sqrt n \int d^D x \langle\langle B |\sum_{n=0}^\infty
C_n^- (\alpha_n^\mu + \tilde\alpha_n^\mu)|\Phi\rangle\rangle \hat A_\mu(x)
, \cr} \eqn\lagint$$
(where $\alpha_0=\tilde \alpha_0=k/2$)
so that  the coupling to massless closed-string states is simply $g\sqrt n \hat
A^\mu
\partial_\mu D$.
[In the light-cone-like frame $\hat A^\mu$ couples to a gauge-invariant
combination of massless fields,
$g\sqrt n \int d^Dx  \left(\hat A_\mu \left(\eta^\mu + \tilde \eta^\mu
\right.\right. $
$ \left.\left. +
ik_\nu h^{\mu\nu} + ik_\nu
h^{\nu\mu} - i k^\mu (\phi + \tilde \phi)\right)\right. $ $\left.
+i \hat A_\mu k^\mu D  \right) $.
Together with an $O(g^2)$ mass term for $\hat A^\mu$, the first term
 can be absorbed
into a redefinition of the $\eta$ and $\eta'$ terms in $S^{\prime\prime}$ which
are then
integrated, leaving
the previous effective interaction between $\hat A^\mu$ and the dilaton.]

Adding \lagint\ to the free action and integrating over $\hat A^\mu$ leads
to the constraint (using $\langle\langle B|B_0^-=\langle\langle B|C_0^+ = 0$),
$$k^\mu \langle\langle B|B_0^+ C_0^-C_0^+|\Phi\rangle\rangle = 0
.\eqn\condition$$
The components of a general closed-string field that enter \condition\
have the expansion in terms of Fock space component states,
$$C_0^-C_0^+|\Phi\rangle\rangle  = \sum_{\sum  n_r^M   =
\sum  \bar n_r^M  } \phi_{\{\bar n_r^M\}}^{\{n_r^M\}} (x) |{\{ n_r^M\}} {\{\bar
n_r^M\}}
\rangle\rangle \otimes |\uparrow\uparrow \rangle\rangle, \eqn\phiexp$$
where ${\{ n_r^M\}} ,{\{\bar n_r^M\}}$ are occupation numbers for the left and
right
moving Fock spaces and the  non-zero modes, $\alpha_r^\mu$, $B_r$ and $C_r $
have been combined
into a
vector of $OSP(1, D-1|2)$ labelled by the index  $M$ (and likewise for
$\tilde\alpha_r^\mu$,
$\tilde B_r$ and $\tilde C_r$).
The condition \condition\ is a constraint on the graded sum (counting states
with an odd number of
right-moving fermionic excitations with a minus sign)  of those component
fields for which there
are equal numbers of left-moving and right-moving Fock space excitations, \ie\
fields of the form
$\phi_{\{n_r^M\}}^{\{n_r^M\}} (x) $.  The combination of component fields at
level $N$ that is
affected by the constraint is
$\phi_N \equiv \sum_{\{n_r^M |  \sum n_r^M = N\}} \phi_{\{n_r^M\}}^{\{n_r^M\}}
(x) /\sqrt{d_N}$,
where the normalization factor $d_N = \int_0^{2\pi}  d\theta
\prod_{n=1}^\infty (1- e^{in\theta })^{-24} e^{-iN\theta}/2\pi$ (which
is equal to the usual degeneracy of open-string states) has been chosen
to give the kinetic term
$$\int d^D x \sum_{N, P=0}^\infty \half \phi_N
D_{NP}\phi_P,\eqn\freefield$$
with $D_{NP} = \delta_{NP} (k^2 + 4N -4)$.
The constraint takes the form
 $k^\mu \sum_N \phi_N(x) \sqrt {d_N} =0.$
Naively, this may be argued to eliminate the non-constant
part of the dilaton field ($D\equiv \phi_1$) since in the
 low-momentum limit closed-string states of
non-zero mass decouple ($\phi_N\sim \sqrt {\alpha'}$ for $N\ne 1$).
The constraint then becomes simply
$k^\mu D=0$ and the dilaton state disappears.

This may be made more explicit by  substituting  \condition\ into the free
closed-string action by simply making
the replacement  $k^\mu D \to -k^\mu \sum' \phi_N \sqrt{d_N}$
in  \freefield,   where  the prime
indicates that the $N=1$ state is missing from the sum
(if the constraint  were used to to
eliminate any other component field an equivalent, but more
complicated,
non-local action would result).   This gives a  modified action that can then
be written as
$$ \int d^D x  {\sum}^{\prime} \half \phi_N M_{NP} \phi_P.
\eqn\mateqn$$
The matrix
M is defined by $M_{NP}= D_{NP} + k^2 v_N v_P$, where
$v_N= \sqrt {d_N}$.
The eigenvalues of $M$, which determine the diagonal form of
a modified free-field action,
are determined by the zeroes of
$$\eqalign{{\det}^{\prime}(M - \lambda I) =&
{\det}^{\prime}
(D-\lambda I)  \left(1 + k^2 (D- \lambda I)_{NP}^{-1}v_Nv_P\right) \cr
       = &{\det}^{\prime} (D-\lambda I)  \left(1 +  k^2
  {\sum}^{\prime}_{N }
{d_{N} \over k^2+4 N - 4 -\lambda}
\right).\cr}\eqn\determine$$
The   zeroes of $\det' (D-\lambda I)$ at $\lambda=k^2+4 N - 4 $ are cancelled
by the poles
of the   sum in parentheses and the eigenvalues of M
are  therefore determined by the zeroes of the last factor.
The masses of particle states are given by the  zero  eigenvalues, \ie\
by the values of $k^2$
at which both ${\det}^{\prime}(M - \lambda I)=0$ and $\lambda=0$.
Since the sum in parentheses in \determine\ is  divergent due to
the contribution of the states with large $N$ we shall proceed by regularizing
it
by  replacing the last factor in \determine\ (setting $\lambda=0$) by
the small-$\epsilon$ limit of the integral representation,
$$\eqalign{\sum_{N=0 }^\infty d_N  {  k^2 e^{-( k^2 +4 N  - 4 )
\epsilon/2}\over
 k^2+ 4N  - 4} =&
k^2\int_\epsilon^\infty{ dl \over 2}e^{(4-  k^2)l/2}
\prod_{n=1}^\infty (1- e^{-2nl})^{-24}\cr
      = & {\pi\over 2}(2\pi)^{13}k^2 \int_0^{1/\epsilon}{dl'\over l^{\prime
14}}e^{l'-\pi^2k^2/l'} \prod_{n=1}^\infty (1-e^{-nl'})^{-24}
\cr}\eqn\intrep$$
(where $l' =2\pi^2/l$) which can be viewed as a functional integral on a
cylinder of minimum length
$\epsilon$.  The divergence of the series arises from the upper limit of the
last
expression as $\epsilon\to 0$, which illustrates that it can be blamed on the
existence of a tachyon
state in the open-string annulus that is dual to the cylinder.

At $k^2=0$ the regulated sum is given by the $N=1$ term which is non-zero so
that there is no massless scalar state, as anticipated.
In fact, there is exactly one zero
between each pair of neighboring poles (all $d_N$ are positive).
However, the divergence of the unregulated sum confuses the issue.
As the
regulator is removed ($\epsilon \to 0$) the far away poles
contribute a positive infinite constant which causes each of
the zeroes to move down towards the adjacent pole with the
lower value of
$\lambda$, approaching the pole as $e^{-1/\epsilon}$.  The
net result is
that the zeroes of the  determinant approach
the same positions as the zeroes of $\det D$
so that the
diagonalised kinetic terms approach those of the usual free-field theory,
including a scalar with
$($mass$)^2$ $\sim e^{-1/\epsilon}$.

This discussion of the constraints has singled out the Lagrange multiplier
component
of the open-string field, thereby accounting for the
divergences of the separate perturbation
theory diagrams.
The effect of the coupling of the infinite number of other
components of the open-string
field  -- auxiliary fields --  to the boundary must also be
taken into account in order to
understand the closed-string spectrum in any detail.

\vskip 0.3cm
\noindent{\it Comments.}
\par\vskip 0.1cm\nobreak
This paper has discussed the ocurrence of constraints on
closed-string fields induced by their
coupling to fixed end-point open-string fields via Dirichlet
world-sheet boundaries.  These
constraints are not taken into account in string
perturbation theory which is consequently plagued
by divergences that arise from the region of moduli space in which the
intermediate open string
in fig.1(c) becomes infinitely long. The solution of the constraints mixes
terms
arising at  all orders in perturbation
theory which makes it difficult to estimate the details of
the closed-string spectrum.  However, the elimination of the dilaton is a
feature that can be argued
for in the low-energy limit provided a world-sheet regulator is introduced.
Without such a
regulator the situation is obscured by a divergence that originates from the
presence of
the \lq tachyon' state of the fixed end-point open string.
If the  dilaton field is indeed  constrained to be constant then
there is no longer a  divergence due to  massless states in
the cylinder coupling to the boundary
state  (recalling that  the dilaton is the only massless state to
which the boundary state couples in the
Dirichlet theory).

Apart from leading to a modification of the massless spectrum, Dirichlet
boundaries are also intimately connected
with a modification of the short-distance properties of the
theory, as noted in earlier work.  For example,  in the presence of a single
Dirichlet
world-sheet boundary  the
fixed-angle high-energy  cross section  decreases  as a  power of the energy,
in contrast to the exponential
decrease characteristic of conventional
closed-string theories and theories with Neumann boundaries.
Furthermore, this power behaviour arises from the region of moduli space in
which the vertex
operators in fig.1(a) approach the boundary  --- a region  that is larger than
(but includes)  the region that gives rise to the open-string divergences.  For
this reason
 the elimination of these divergences by solving the constraints
on the closed-string field should not destroy the power behaviour of
fixed-angle scattering.
Higher order effects due to the insertion of multiple Dirichlet boundaries
gives rise to
modifications of the leading power behaviour reminiscent of the logartihmic
corrections
to naive scaling behaviour in renormalizable field theory ([\greenb] and
references therein).
Further evidence of modified short-distance properties
is provided by the high-temperature behaviour of the theory.
At finite temperature  the fixed
end-point open string can wind around the compact
temperature direction and  states of  winding $q$ develop an
effective mass proportional to $\beta$ (the inverse
temperature) since  $L_{0D}  = \beta^2 q^2/4\pi^2+ N  $.  In the
Dirichlet theory
such open strings arise as intermediate states in the
closed-string propagator (as in fig.1(c)).  Once again the presence of an $N=1$
vector state
(now with \lq mass' $\beta q /2\pi $)
leads to  an important  temperature-dependent
renormalization
of the closed-string spectrum as $\beta\to 0$
\REF\greenc{M.B. Green, {\it Temperature Dependence of
String
Theory in the Presence of World-Sheet Boundaries}, Phys.
Lett.
{\bf 282B}  (1992) 380 .}  [\greenc].  This makes contact
with the expression deduced in the large-$n$ limit of $U(n)$ Yang--Mills theory
by
Polchinski\REF\polchinew{J. Polchinski, {\it High
temperature limit of the confining phase},
Phys. Rev. Lett. {\bf 68}   (1992) 1267.} [\polchinew]

The possibility that the presence of point-like structure may be  related
directly to
the massless spectrum of the theory is most intriguing.  Alas, as yet there is
no evidence that
the massless graviton (that seems to be an intrinsic feature of usual string
theory) is absent
in the Dirichlet theory as it would be in any  realistic description of
large-$n$ Yang--Mills theory.

The simplest bosonic theory discussed here is ill-defined due to the presence
of
tachyonic states that indicate  instabilities   requiring  a more sophisticated
treatment  which  would take into account a condensate of Dirichlet boundaries
as well
as handles.  It might be that a supersymmetric theory such as the one
outlined in \REF\greenv{M.B.  Green,
{\it Wilson--Polyakov loops for critical strings and
superstrings at finite temperature}, Nucl.  Phys. {\bf B381}
(1992) 201.}
[\greenv] could form the basis of a more consistent,
tachyon-free theory.   Another possible arena for these ideas is in the context
of sub-critical
string theory -- indeed, the usual arguments concerning the critical dimension
do not apply in
an obvious manner since the constraints mix different orders in the string
perturbation theory.
Generalizations of the boundary conditions are also possible.   For example,
a theory with both Neumann and Dirichlet conditions describes open strings
with   end-points
which may be  fixed or free.  Another possibly important variant is a theory
which involves
a sum over all possible ways of dividing each  boundary
into segments  with Neumann or constant Dirichlet conditions where the
target-space position
and the length of each
Dirichlet segment is integrated [\greenb] (which may make use of ideas in
\REF\schwarza{J.H. Schwarz,  {\it Off-shell dual amplitudes without ghosts},
{Nucl. Phys.} {\bf B65} (1973) 131.}\REF\fairliea{E.F. Corrigan and
D.B. Fairlie, {\it Off-shell states in dual resonance theory}, Nucl. Phys. {\bf
B91} (1975)
527.}\REF\greenv{M.B. Green, {\it Locality and currents for the dual string},
Nucl. Phys.{\bf B103} (1976) 333.} [\schwarza-\greenv]).  This introduces
point-like structure
at the string end-points.
More general backgrounds than the Minkowski space considered in this paper may
be
introduced by considerations of  boundary conditions in conformal field theory
based on  \REF\cardya{J.L.  Cardy,{\it Boundary conditions
in conformal field theory},
Nucl.  Phys. {\bf B324} (1989)  581.} [\cardya].

\smallskip
\centerline{Acknowledgments}
I am grateful to Ed Corrigan, Chris Hull and other visitors to the Isaac Newton
Institute for useful discussions.

\refout

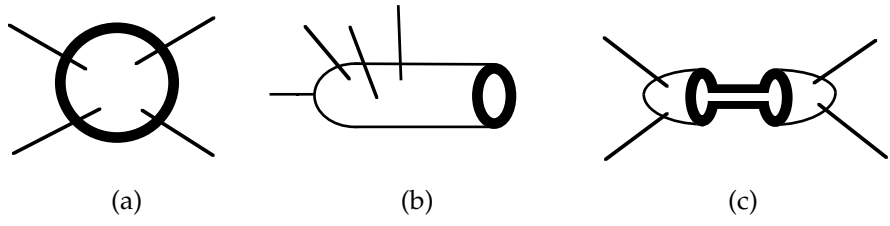
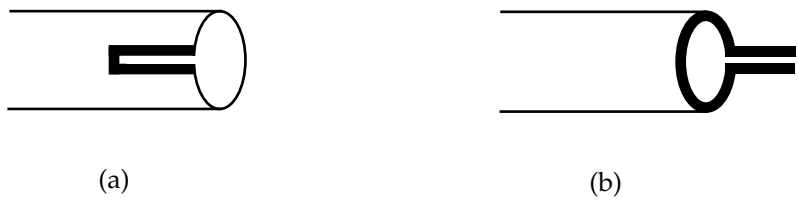
\bye